\documentclass[a4paper,11pt]{article}
\usepackage[hmargin=3cm,vmargin=3cm]{geometry,}
\usepackage{graphicx, amssymb, textcomp, hyperref, amsmath}
\numberwithin{equation}{section}
\title{\bf Novel quasi-exactly solvable models with anharmonic singular potentials}
\author{\Large Davids Agboola and Yao-Zhong Zhang}
\date{\it School of Mathematics and Physics, The University of Queensland, \\Brisbane, QLD 4072, Australia}
\begin{document}
\maketitle
\vspace{0.5in}
\noindent {\bf Abstract:} We present new quasi-exactly solvable models with
inverse quartic, sextic, octic and decatic power potentials, respectively. We solve these models exactly by means of
the functional Bethe ansatz method. For each case, we give closed-form solutions for the energies and the wave functions
as well as analytical expressions for the allowed potential parameters in terms of the roots of a set of algebraic equations.

\vspace{.3in}


\vskip.5in
\section{Introduction}
Since the early works on singular potentials (see e.g. \cite{Plesset1932}-\cite{VW1954}),
an extensive literature has been developed on the subject.
By singular potentials, we mean those potentials
$V(r)$ with property
$\lim_{r\rightarrow 0}r^2V(r)\rightarrow\infty,$
although sometimes the inverse-square potential is also regarded as singular.
The investigation of singular potentials covers a wide range of physical and mathematical interest.
In view of the availability of a comprehensive review article by Frank, Land and Spector \cite{FLS1971}
on singular potentials and applications, here we will only refer to the main points of the topic.

One of the early works that generated much interest in the study of singular potentials was the one by
Predazzi and Regge \cite{PR1962}, who argued that real world interactions were likely to be highly singular and thus
the study of singular potentials rather than regular potentials might be more relevant physically.
This was thereafter followed by the applications of the singular potentials $r^{-n}$ $(n>2)$ in the study of
the (p, p) and (p, $\pi$) processes in high energy physics \cite{Tiktopoulous1965,Kouris1966}. The
interactions of nucleons with K-mesons and $\alpha$-$\alpha$ scattering have been reproduced by repulsive
singular potentials \cite{SC1968}. As examples from non-relativistic quantum mechanics,
the problem of high-energy scattering by strongly singular potentials was investigated by many authors
(see e.g. \cite{PW1964,Dolinszky1980,Esposito1998}). In molecular physics, inter-atomic or intermolecular forces are mostly represented
by singular potentials whose parameters are determined phenomenologically. Examples are the Mie-type potential
\cite{IK2007,ES1984} and the Lennard-Jones potential $r^{-12}$ \cite{DV1987} which describe the interaction between
diatomic molecules and two non-polar molecules, respectively.

In field theory, the importance of singular potentials emerged from the efforts to find effective
potentials of field-theoretic interactions in the Bethe-Salpeter equations.
In \cite{BBFFT1963}, Bastai, Bertocchi, Fubini, Furlan and Tonin  discovered a remarkable relationship
between the renormalizability of a field theory and the regularity of the effective potential.
That is, the effective potentials for non-renormalizable field theories are singular,
whereas superrenormalizable and renormalizable field theories give rive to regular and "transition" potentials, respectively.
Thus, any new insight gained in the analysis of singular potentials could lead to a better
understanding of quantum field theories which are not perturbatively renormalizable.

In this paper, we consider a class of most frequently discussed singular potentials
in nonrelativistic quantum mechanics, that is the spherically symmetric, inverse power potentials of the form
\cite{FLS1971,BJ1983,MRSW1987},
$$
V(r)=\sum_{k=0}^N\frac{\lambda_k}{r^{\alpha_k}},
$$
where  $r\in (0,\infty)$, $\alpha_k$ and $\lambda_k$ are positive real numbers.
The corresponding Schr\"odinger equation for the radial wave function $\Psi(r)$ is given by
$$
\left[-\frac{d^2}{dr^2}+\frac{\ell(\ell+1)}{r^2}+\omega^2r^2+2V(r)\right]\Psi(r)=2E\Psi(r),~~~~\omega\geq 0,
$$
where $\ell=-1, 0, 1,\cdots$, and $E$ is the energy eigenvalue. Much of the investigation of
singular potentials is concerned with the solutions of this equation.
Several approximation techniques are
avaliable in the literature for calculating the eigenvalues and wave functions of some
inverse-power potentials (see e.g. \cite{DK1975,Harrell1977,Znojil1989-90,SHv2003,CHS2008,SD2010}).

We show that at least four classes of singular potentials in the above family
are quasi-exactly solvable, i.e. have polynomial solutions \cite{Turbiner96,GKO93,Ushveridze94}.
Namely, we present new quasi-exactly solvable models with 
inverse quartic, sextic, octic and decatic power potentials, respectively.
We solve these models exactly by using the functional Bethe ansatz method presented in \cite{Zhang11}.
For each model, we obtain closed-form solutions for the energies and wave functions as well as analytic expressions for
the allowed potential parameters in terms of the roots of a set of algebraic (Bethe ansatz) equations.
To our best knowledge, our results on the singular quartic and sextic power potentials
are largely new, while our results on the inverse octic and decatic power potentials are completely new.

\section{Quasi-exactly solvable inverse quartic power potential}
The inverse quartic power potential
\begin{equation}\label{eq:2.1}
V(r)=\frac{a}{r}+\frac{b}{r^2}+\frac{c}{r^3}+\frac{d}{r^4},~~~~d>0
\end{equation}
was investigated by Predazzi and Regge \cite{PR1962} to determine analytic properties of the scattering amplitude
in the case of a singular potential. The corresponding radial Schr\"odinger equation is given by
\begin{equation}\label{eq:2.2}
\left[-\frac{d^2}{dr^2}+\frac{\ell(\ell+1)
}{r^2}+\omega^2r^2+\frac{2a}{r}+\frac{2b}{r^2}+\frac{2c}{r^3}+\frac{2d}{r^4}\right]\Psi(r)=2E\Psi(r).
\end{equation}
Phenomenologically, this type of singular potential is a very useful form of anharmonicity in physical
applications  \cite{Znojil1989-90}.

Here we are interested in finding exact solutions of the Schr\"odinger equation.
To this end,  we first extract the appropriate asymptotic behaviour of the wave function $\Psi(r)$ by making a substitution.
After a brief inspection of the differential equation, we arrive at the transformation,
\begin{equation}\label{quartic-transformation}
\Psi(r)=r^\gamma\,\exp\left[Ar^2+Br+\frac{D}{r}\right]x(r),
\end{equation}
where $\gamma, A, B$ and $D$ are constant parameters to be determined.  Substituting into the differential equation, we
find that acceptable asymptotic behaviour of $\Psi(r)$ requires
\begin{equation}
\gamma=1+\frac{c}{\sqrt{2d}}>0,~~~~A=-\frac{\omega}{2},~~~~D=-\sqrt{2d}.
\end{equation}
At this stage, the parameter $B$ is free.
Now the differential equation for $x(r)$ reads 
\begin{equation}\label{eq:2.4}
r^2x''(r)+2\left[-\omega r^3+Br^2+\left(1+\frac{c}{\sqrt{2d}}\right)r+\sqrt{2d}\right]x'(r)
$$$$+\left\{-2\omega Br^3+\left[2E+B^2-\omega\left(3+\frac{2c}{\sqrt{2d}}\right)\right]r^2\right.
$$$$\left.+2\left[B\left(1+\frac{c}{\sqrt{2d}}\right)-a-\omega\sqrt{2d}\right]r\right\}x(r)=\left(2b+\ell(\ell+1)-4B\sqrt{2d}\right)x(r).
\end{equation}

In the following,  we show that if $B=0$ or $\omega=0$, then Eq.\,\eqref{eq:2.4} is quasi-exactly solvable provided that
the potential parameters satisfy certain constraints, and exact solutions are given by the degree $n$ polynomials
\begin{equation}\label{quartic-wavefunction}
x(r)=\prod_{i=1}^n(r-r_i),\hspace{0.3in}x(r)\equiv 1\hspace{0.1in}\mbox{for}\hspace{0.1in} n=0.
\end{equation}

\vskip.2in
\noindent{\bf\large Case 1: $B=0$}
\vskip.1in
In this case, Eq. \eqref{eq:2.4} becomes
\begin{equation}\label{eq:3.10}
r^2x''(r)+2\left[-\omega r^3+\left(1+\frac{c}{\sqrt{2d}}\right)r+\sqrt{2d}\right]x'(r)+\hspace{1in}
$$
$$
\left\{\left[2E-\omega\left(3+\frac{2c}{\sqrt{2d}}\right)\right]r^2-2\left[a+\omega\sqrt{2d}\right]r\right\}x(r)=\left[2b+\ell(\ell+1)\right]x(r).
\end{equation}
Substituting \eqref{quartic-wavefunction} into \eqref{eq:3.10} and applying
the functional Bethe ansatz method in \cite{Zhang11} which is outlined in the
Appendix, we obtain the closed-form expressions for the energies and the wave functions
\begin{equation}\label{eq:3.11}
E_n=\omega\left(n+\frac{3}{2}+\frac{c}{\sqrt{2d}}\right),$$$$
\Psi_n(r)=r^{1+c/\sqrt{2d}}\left[\prod_{i=1}^n(r-r_i)\right]\exp\left(-\frac{\omega}{2}r^2-\frac{\sqrt{2d}}{r}\right),
\end{equation}
and the constraints for the potential parameters
\begin{equation}\label{eq:3.12}
a=-\omega\left(\sqrt{2d}+\sum_{i=1}^nr_i\right),
\end{equation}
\begin{equation}\label{eq:3.13}
n\left(n+1+\frac{2c}{\sqrt{2d}}\right)-2\omega\sum_{i=1}^nr_i^2=2b+\ell(\ell+1),
\end{equation}
where the roots $\{r_i\}$ satisfy the Bethe ansatz equations
\begin{equation}\label{eq:3.14}
\sum_{j\neq i}^n\frac{1}{r_i-r_j}=\frac{\omega r^3_i-\left(1+\frac{c}{\sqrt{2d}}\right)r_i-\sqrt{2d}}{r^2_i},\hspace{0.5in}i=1,2,\dots,n.
\end{equation}
The wave functions $\Psi_n(r)$ are squarely integrable, i.e. $\int^\infty_0\,|\Psi_n(r)|^2\,dr<\infty$. In fact the
normalization constants of  $\Psi_n(r)$ can be calculated analytically by means of the standard integral
$$
\int^\infty_0\, r^\nu\exp\left(-\mu_1 r^2-\frac{\mu_2}{r}\right)dr=\frac{\mu_2^\nu}{2^{\nu+1}\sqrt{\pi\mu_1}}\,
G_{02}^{10}\left(\frac{1}{2},\frac{1}{2}-\frac{\nu}{2},-\frac{\nu}{2}\left|~\frac{1}{4}\mu_1\mu_2^2\right.\right)
$$
for Re$(\nu)>0$, Re$(\mu_1)>0$ and Re$(\mu_2)>0$, where $G_{pq}^{mn}$ is the Meijer G-function.

As examples of the above general expressions of the exact solutions,
we study the ground and first excited states of the system in some detail.
It is easy to see that $x=1$ is a solution of \eqref{eq:3.10} for certain values of the potential parameters.
This solution corresponds to the $n=0$ case in the general expressions above. Indeed, from Eqs. \eqref{eq:3.11}
and \eqref{eq:3.12}, \eqref{eq:3.13}, we obtain
\begin{equation}\label{eq:3.15}
E_0=\omega\left(\frac{3}{2}+\frac{c}{\sqrt{2d}}\right),$$$$
a=-\omega\sqrt{2d}\hspace{0.2in}\mbox{and}\hspace{0.2in}2b=-\ell(\ell+1).
\end{equation}
The corresponding wave function is
\begin{equation}\label{eq:3.16}
\Psi_0(r)=r^{1+c/\sqrt{2d}}\exp\left(-\frac{\omega}{2}r^2-\frac{\sqrt{2d}}{r}\right).
\end{equation}
This wave function has no nodes and so the state described by it is the ground state of the system.

The first excited state solution of the system corresponds to the $n=1$ case of the general expressions above.
Explicitly, the energy and wave function read, respectively,
\begin{equation}\label{eq:3.17}
E_1=\omega\left(\frac{5}{2}+\frac{c}{\sqrt{2d}}\right),$$$$
\Psi_1(r)=r^{1+c/\sqrt{2d}}(r-r_1)\exp\left(-\frac{\omega}{2}r^2-\frac{\sqrt{2d}}{r}\right)
\end{equation}
with the constraints for the potential parameters,
\begin{equation}\label{eq:3.18}
a=-\omega\left(r_1+\sqrt{2d}\right),$$$$
b=\left(1+\frac{c}{\sqrt{2d}}\right)-\omega r^2_1-\frac{\ell(\ell+1)}{2}.
\end{equation}
Here the root $r_1$ is determined by the Bethe ansatz equation
$$\omega r_1^2-\left(1+\frac{c}{\sqrt{2d}}\right)r_1-\sqrt{2d}=0,$$
giving rise to
\begin{equation}\label{eq:3.19}
r_1=\frac{1}{2\omega}\left[\left(1+\frac{c}{\sqrt{2d}}\right)\pm\sqrt{\left(1+\frac{c}{\sqrt{2d}}\right)^2+4\omega\sqrt{2d}}\right].
\end{equation}

\vskip.2in
\noindent{\bf\large Case 2: $\omega=0$}
\vskip.1in
As $A=0$ in this case, in order for the wave function $\Psi(r)$ to have acceptable asymptotic behaviour,
we require the constant parameter $B$ to be negative.

The radial Schr\"odinger equation corresponding to this case was previously investigated in \cite{OS91}
(see also \cite{DM99} for the ground state solution of the Schr\"odinger equation). Here we
solve this system exactly by using the functional Bethe ansatz method of \cite{Zhang11}.

If we set
\begin{equation}
2E+B^2=0,
\end{equation}
then Eq.\,\eqref{eq:2.4} becomes
\begin{equation}\label{eq:3.20}
r^2x''(r)+2\left[Br^2+\left(1+\frac{c}{\sqrt{2d}}\right)r+\sqrt{2d}\right]x'(r)\hspace{0.5in}$$$$
+2\left[B\left(1+\frac{c}{\sqrt{2d}}\right)-a\right]r\,x(r)=\left[2b+\ell(\ell+1)-4B\sqrt{2d}\right]x(r).
\end{equation}
Substituting \eqref{quartic-wavefunction} into \eqref{eq:3.20} and following the procedure in \cite{Zhang11}
(or applying the results of the Appendix), we obtain the relations,
\begin{equation}\label{eq:3.21}
B=\frac{a}{n+1+\frac{c}{\sqrt{2d}}},
\end{equation}
\begin{equation}\label{eq:3.23}
2B\left(2\sqrt{2d}+\sum_{i=1}^nr_i\right)+n\left(n+1+\frac{2c}{\sqrt{2d}}\right)=2b+\ell(\ell+1),
\end{equation}
and the Bethe ansatz equations,
\begin{equation}\label{eq:3.24}
\sum_{j\neq i}^n\frac{1}{r_i-r_j}+\frac{Br^2_i+\left(1+\frac{c}{\sqrt{2d}}\right)r_i+\sqrt{2d}}{r^2_i}=0\hspace{0.5in}i=1,2,\dots,n.
\end{equation}

As mentioned above, we require $B<0$. It follows from \eqref{eq:3.21} that the potential parameter $a$ has to be negative. So
we assume that $a<0$ if $\omega=0$.

Thus we have the closed form expressions for the energies and the wave functions,
\begin{equation}
E_n=-\frac{1}{2}\left(\frac{a}{n+1+\frac{c}{\sqrt{2d}}}\right)^2,$$$$
\Psi_n(r)=r^{1+c/\sqrt{2d}}\left[\prod_{i=1}^n(r-r_i)\right]\exp\left(\frac{a}{n+1+\frac{c}{\sqrt{2d}}}r-\frac{\sqrt{2d}}{r}\right),
\end{equation}
and the constraint for the potential parameters
\begin{equation}
2a\left(2\sqrt{2d}+\sum_{i=1}^nr_i\right)+n\left(n+1+\frac{2c}{\sqrt{2d}}\right)\left(n+1+\frac{c}{\sqrt{2d}}\right)
$$$$\hspace{1.5in}=\left(n+1+\frac{c}{\sqrt{2d}}\right)\left[2b+\ell(\ell+1)\right]
\end{equation}
with the roots $\{r_i\}$ determined by \eqref{eq:3.24}. Again the wave functions $\Psi_n(r)$ are squarely integrable,
and the normalization constants  can be computed analytically with the help of the standard integral \cite{GR65}
$$
\int^\infty_0\, r^\nu\exp\left(-\mu_1 r-\frac{\mu_2}{r}\right)dr
=2\left(\frac{\mu_2}{\mu_1}\right)^{(\nu+1)/2}\,{\rm BesselK}\left(\nu+1,2\sqrt{\mu_1\mu_2}\right)
$$
for Re$(\nu)>0$, Re$(\mu_1)>0$ and Re$(\mu_2)>0$.

Similar to the Case 1 above, the ground state energy and wave function are obtained from
the general expressions by setting $n=0$,
\begin{equation}\label{eq:3.25}
E_0=-\frac{1}{2}\left(\frac{a}{1+\frac{c}{\sqrt{2d}}}\right)^2,
$$$$
\Psi_0(r)=r^{1+c/\sqrt{2d}}\exp\left(\frac{a\sqrt{2d}}{c+\sqrt{2d}}r-\frac{\sqrt{2d}}{r}\right)
\end{equation}
with the constraint for
\begin{equation}\label{eq:3.26}
c=\frac{8ad}{2b+\ell(\ell+1)}-\sqrt{2d}.
\end{equation}
The first excited state solution corresponds to the $n=1$ case with the energy and the wave function given by
\begin{equation}\label{eq:3.27}
E_1=-\frac{1}{2}\left(\frac{a}{2+\frac{c}{\sqrt{2d}}}\right)^2,$$$$
\Psi_1(r)=r^{1+c/\sqrt{2d}}(r-r_1)\exp\left(\frac{a\sqrt{2d}}{c+2\sqrt{2d}}r-\frac{\sqrt{2d}}{r}\right),
\end{equation}
where the potential parameters satisfy the constraint
\begin{equation}\label{eq:3.28}
2c^2-c\sqrt{2d}\left[2b-6+\ell(\ell+1)\right]+4d\left[a\left(r_1+2\sqrt{2d}\right)-2(b-1)-\ell(\ell+1)\right]=0
\end{equation}
and the root $r_1$ is determined by
\begin{equation}\label{eq:3.29}
\frac{a\, r_1^2}{2+\frac{c}{\sqrt{2d}}}+\left(1+\frac{c}{\sqrt{2d}}\right)r_1+\sqrt{2d}=0$$$$\Rightarrow\hspace{0.2in}
r_1=\frac{\left(c+2\sqrt{2d}\right)}{2a\sqrt{2d}}\left[-\left(1+\frac{c}{\sqrt{2d}}\right)
\pm\sqrt{\left(1+\frac{c}{\sqrt{2d}}\right)^2-\frac{8ad}{c+2\sqrt{2d}}}\right].
\end{equation}

\section{Quasi-exactly solvable inverse sextic power potential}
Pais and Wu \cite{PW1964} studied the problem of scattering by the singular potential
$d/ r^{2+2n}+e/ r^{2+n}$ $(n>1)$ in non-relativistic quantum mechanics.
In this section we will consider the $n=2$ case, i.e. the  inverse sextic power potential \cite{Znojil1989-90}
\begin{equation}\label{eq:2.5}
V(r)=\frac{e}{r^4}+\frac{d}{r^6},~~~~d>0.
\end{equation}
This potential has been used in atomic, molecular and nuclear physics \cite{BJ1983,MRSW1987}.
The corresponding radial Schr\"odinger equation is
\begin{equation}\label{eq:2.6}
\left[-\frac{d^2}{dr^2}+\frac{\ell(\ell+1)}{r^2}+\omega^2r^2+\frac{2e}{r^4}+\frac{2d}{r^6}\right]\Psi(r)=2E\Psi(r).
\end{equation}
The exact ground state solution of the Schr\"odinger equation was previously obtained in \cite{Kaushal91}. Here we
review some of the results given in \cite{AZ11} and present the general exact solutions of the system.

In order to find the exact solutions of the Schr\"odinger equation, similar to the inverse quartic power potential
case we extract the appropriate asymptotic behaviour of the wave function $\Psi(r)$ by making the substitution,
\begin{equation}\label{sextic-Trans}
\Psi(r)=r^{3/2+e/\sqrt{2d}}\,\exp\left[-\frac{\omega}{2}r^2-\frac{\sqrt{2d}}{2}\frac{1}{r^2}\right]v(r),~~~~3/2+e/\sqrt{2d}>0,
\end{equation}
we obtain the differential for $v(r)$,
\begin{equation}\label{sextic-Eq2}
v''(r)+\frac{2}{r}\left(-\omega r^2+\frac{3}{2}+\frac{e}{\sqrt{2d}}+\frac{\sqrt{2d}}{r^2}\right)v'(r)
+2\left[E-\omega\left(2+\frac{e}{\sqrt{2d}}\right)\right]v(r)
$$
$$
=\frac{1}{r^2}\left[2\omega \sqrt{2d}+\left(\ell+\frac{1}{2}\right)^2-\left(\frac{e}{\sqrt{2d}}+1\right)^2\right]v(r)
\end{equation}
Then the change of variable $t=r^2$ transforms \eqref{sextic-Eq2} into the form,
\begin{equation}\label{sextic-Eq3}
t^2v''(t)+\left[-\omega t^2+\left(2+\frac{e}{\sqrt{2d}}\right)t+\sqrt{2d}\right]v'(t)
+\frac{t}{2}\left[E-\omega\left(2+\frac{e}{\sqrt{2d}}\right)\right]v(t)
$$
$$
=\frac{1}{4}\left[2\omega \sqrt{2d}+\left(\ell+\frac{1}{2}\right)^2-\left(\frac{e}{\sqrt{2d}}+1\right)^2\right]v(t).
\end{equation}

This equation has the degree $n$ polynomial solutions
\begin{equation}\label{sextic-wavefunction}
v(t)=\prod_{i=1}^n(t-t_i),\hspace{0.3in}v(t)\equiv 1\hspace{0.1in}\mbox{for}\hspace{0.1in} n=0
\end{equation}
with distinct roots $t_i$ provided that the potential parameters satisfy certain constraints \cite{AZ11}.
The closed-form expressions for the energies and the wave functions are
\begin{equation}\label{sextic-E}
E_n=\omega\left(2n+2+\frac{e}{\sqrt{2d}}\right),$$$$
\Psi_n(r)=r^{3/2+e/\sqrt{2d}}\left[\prod_{i=1}^n(r^2-t_i)\right]\,\exp\left[-\frac{\omega}{2}r^2-\frac{\sqrt{2d}}{2}\frac{1}{r^2}\right]
\end{equation}
and the constraint for the potential parameters reads
\begin{equation}\label{sextic-Constraint}
2\omega\left(\sqrt{2d}+2\sum_{i=1}^nt_i\right)+\left(\ell+\frac{1}{2}\right)^2
=4n\left(n+1+\frac{e}{\sqrt{2d}}\right)+\left(\frac{e}{\sqrt{2d}}+1\right)^2,
\end{equation}
while the roots $\{t_i\}$ are determined by the Bethe ansatz equations,
\begin{equation}\label{sextic-BAEs}
\sum_{j\neq i}^n\frac{2}{t_i-t_j}=\frac{\omega t_i^2-\left(2+\frac{e}{\sqrt{2d}}\right)t_i-\sqrt{2d}}{t_i^2},~~~~i=1,2,\cdots,n.
\end{equation}
The wave functions $\Psi_n(r)$ are squarely integrable, and the normalization constants can be evaluated by using the
standard integral \cite{GR65}
$$
\int^\infty_0\, r^\nu\exp\left(-\mu_1 r^2-\frac{\mu_2}{r^2}\right)dr
=\left(\frac{\mu_2}{\mu_1}\right)^{(\nu+1)/4}\,{\rm BesselK}\left(\frac{\nu+1}{2},2\sqrt{\mu_1\mu_2}\right)
$$
for Re$(\nu)>0$, Re$(\mu_1)>0$ and Re$(\mu_2)>0$.

As examples of the above general expressions for the exact solutions, we study the ground and first excited states of the system.
The $n=0$ case gives the ground state energy and wave function
\begin{equation}\label{sextic-E0}
E_0=\omega\left(2+\frac{e}{\sqrt{2d}}\right),$$$$
\Psi_0(r)= r^{3/2+e/\sqrt{2d}}\,\exp\left[-\frac{\omega}{2}r^2-\frac{\sqrt{2d}}{2}\frac{1}{r^2}\right]
\end{equation}
with the potential parameters constrained by
\begin{equation}\label{sextic-Constraint0}
2\omega\sqrt{2d}+\left(\ell+\frac{1}{2}\right)^2=\left(\frac{e}{\sqrt{2d}}+1\right)^2.
\end{equation}

The first excited state solution corresponds to the $n=1$ case of the general expressions
Eqs. \eqref{sextic-Trans} and \eqref{sextic-wavefunction}-\eqref{sextic-BAEs}. The energy and
wave function are given respectively by
\begin{equation}\label{sextic-E1}
E_1=\omega\left(4+\frac{e}{\sqrt{2d}}\right),$$$$
\Psi_1(r) = r^{3/2+e/\sqrt{2d}}\,\left(r^2-t_1\right)\,\exp\left[-\frac{\omega}{2}r^2-\frac{\sqrt{2d}}{2}\frac{1}{r^2}\right],
\end{equation}
where the root $t_1$ is determined by the Bethe ansatz equation,
\begin{equation}\label{sextic-BAEs1}
\omega t_1^2-\left(2+\frac{e}{\sqrt{2d}}\right)t_1-\sqrt{2d}=0
$$
$$\Rightarrow\hspace{0.1in}t_1=\frac{1}{2\omega}\left(2+\frac{e}{\sqrt{2d}}\pm\sqrt{\left(2+\frac{e}{\sqrt{2d}}\right)^2
   +4\omega\sqrt{2d}}\right)
\end{equation}
and  the potential parameters obey the constraint,
\begin{equation}\label{sextic-Constraint1b}
\frac{1}{4}\left[\frac{e^2}{2d}-2\omega\sqrt{2d}+5+\frac{4e}{\sqrt{2d}}-\left(\ell+\frac{1}{2}\right)^2\right]^2
={\left(2+\frac{e}{\sqrt{2d}}\right)^2+4\omega\sqrt{2d}}.
\end{equation}
Here \eqref{sextic-BAEs1} has been used in deriving this equation from \eqref{sextic-Constraint}.

\section{Quasi-exactly solvable inverse octic power potential}
In this section, we consider the inverse octic power potential
\begin{equation}\label{eq:2.10}
V(r)=\frac{a}{r}+\frac{b}{r^2}+\frac{c}{r^3}+\frac{d}{r^4}+\frac{e}{r^5}+\frac{f}{r^6}+\frac{g}{r^7}+\frac{h}{r^8},~~~~h>0.
\end{equation}
This potential is an extension of the inverse quartic and sextic power potentials presented in the previous two sections.
We will show that, in spite of the presence of additional coupling constants, this potential remains to be quasi-exactly solvable.

The corresponding radial Schr\"odinger equation reads
\begin{equation}\label{eq:2.11}
\left[-\frac{d^2}{dr^2}+\frac{\ell(\ell+1)
}{r^2}+\omega^2r^2+\frac{2a}{r}+\frac{2b}{r^2}+\frac{2c}{r^3}+\frac{2d}{r^4}+\frac{2e}{r^5}+\frac{2f}{r^6}+\frac{2g}{r^7}+\frac{2h}{r^8}\right]\Psi(r)=2E\Psi(r).
\end{equation}
To obtain exact solutions, we need to extract the asymptotic behaviour of the wave function $\Psi(r)$ and
transform the differential equation into an appropriate form.
Our experience with the the previous two simpler cases suggests the following substitution for $\Psi(r)$:
\begin{equation}\label{eq:2.12}
\Psi(r)=r^\beta\exp\left[A r^2+Br+\frac{D}{r}+\frac{F}{r^2}+\frac{G}{r^3}\right]y(r),
\end{equation}
where $\beta, A, B, D, F$ and $G$ are as yet unknown constant parameters. Substituting into the Schr\"odinger equation and
after some algebras we find that the acceptable asymptotic behaviour of $\Psi(r)$ requires
\begin{eqnarray}
\beta&=&2+\frac{e}{\sqrt{2h}}+\frac{g}{4}\sqrt{\frac{2}{h^3}}\left(\frac{g^2}{4h}-f\right)>0,\nonumber \\
A&=&-\frac{\omega}{2},~~~~~D=-\frac{1}{\sqrt{2h}}\left(f-\frac{g^2}{4h}\right),\nonumber \\
F&=&-\frac{g}{2\sqrt{2h}},~~~~~G=-\frac{1}{3}\sqrt{2h}. \label{parameter-beta}
\end{eqnarray}
The parameter $B$ is still free at this stage.
Now the differential equation for $y(r)$ is given by
\begin{equation}\label{eq:2.14}
r^4y''(r)+2\left[-\omega r^5+Br^4+\beta r^3+\frac{r^2}{\sqrt{2h}}\left(f-\frac{g^2}{4h}\right)+\frac{gr}{\sqrt{2h}}+\sqrt{2h}\right]y'(r)
$$$$+\left\{-2\omega Br^5+\left(2E+B^2-\omega-2\beta\omega\right)r^4
  -2\left[a-\beta B+\frac{\omega}{\sqrt{2h}}\left(f-\frac{g^2}{4h}\right)\right]r^3\right.
$$$$\left.+\left[(\beta-\ell)(\beta-\ell-1)-2b+\frac{2B}{\sqrt{2h}}\left(f-\frac{g^2}{4h}\right)-\frac{2g\omega}{\sqrt{2h}} \right]r^2\right.
$$$$\left.-2\left[c-\frac{\beta-1}{\sqrt{2h}}\left(f-\frac{g^2}{4h}\right)+\omega \sqrt{2h} -\frac{gB}{\sqrt{2h}}\right]r\right\}y(r)
$$$$\hspace{0.5in}=\left[2d-\frac{g(2\beta-3)}{\sqrt{2h}}-2B\sqrt{2h}-\frac{1}{2h}\left(f-\frac{g^2}{4h}\right)^2\right]y(r).
\end{equation}

If $B=0$ or $\omega=0$, then Eq.\,\eqref{eq:2.14} is quasi-exactly solvable provided that
the potential parameters satisfy certain constraints, and exact solutions are given by the degree $n$ polynomials
\begin{equation}\label{octic-wavefunction}
y(r)=\prod_{i=1}^n(r-r_i),\hspace{0.3in}y(r)\equiv 1\hspace{0.1in}\mbox{for}\hspace{0.1in} n=0.
\end{equation}
This is shown as follows.

\vskip.2in
\noindent{\bf\large Case 1: $B=0$}
\vskip.1in
In this case, equation \eqref{eq:2.14} becomes
\begin{equation}\label{eq:3.38}
r^4y''(r)+2\left[-\omega r^5+\beta r^3+\frac{r^2}{\sqrt{2h}}\left(f-\frac{g^2}{4h}\right)+\frac{gr}{\sqrt{2h}}+\sqrt{2h}\right]y'(r)
$$$$+\left\{\left(2E-\omega-2\beta\omega\right)r^4-2\left[a+\frac{\omega}{\sqrt{2h}}\left(f-\frac{g^2}{4h}\right)\right]r^3\right.
$$$$\left.+\left[(\beta-\ell)(\beta-\ell-1)-2b-\frac{2g\omega}{\sqrt{2h}} \right]r^2
-2\left[c-\frac{\beta-1}{\sqrt{2h}}\left(f-\frac{g^2}{4h}\right)+\omega \sqrt{2h}\right]r\right\}y(r)
$$$$\hspace{0.5in}=\left[2d-\frac{g(2\beta-3)}{\sqrt{2h}}-\frac{1}{2h}\left(f-\frac{g^2}{4h}\right)^2\right]y(r)
\end{equation}
Substituting \eqref{octic-wavefunction} into this equation and applying the results in the Appendix,
we obtain the closed form expressions for the energies and the wave functions
\begin{equation}\label{eq:3.39}
E_n=\omega\left[n+\frac{5}{2}+\frac{e}{\sqrt{2h}}+\frac{g}{4}\sqrt{\frac{2}{h^3}}\left(\frac{g^2}{4h}-f\right)\right],$$$$
\Psi_n(r)=r^\beta \left[\prod_{i=1}^n(r-r_i)\right]\exp\left(-\frac{\omega}{2}r^2-\frac{1}{\sqrt{2h}}
\left(f-\frac{g^2}{4h}\right)\frac{1}{r}-\frac{g}{2\sqrt{2h}}\frac{1}{r^2}-\frac{\sqrt{2h}}{3}\frac{1}{r^3}\right)
\end{equation}
with $\beta$ given by \eqref{parameter-beta} and  the constraints for the potential parameters,
\begin{equation}\label{eq:3.40}
a=-\omega\left[\frac{1}{\sqrt{2h}}\left(f-\frac{g^2}{4h}\right)+\sum_{i=1}^nr_i\right],$$$$
b=\frac{1}{2}\left[(\beta+\ell)(\beta-\ell-1)+n(n+2\beta-1)\right]-\frac{g\omega}{\sqrt{2h}}-\omega\sum_{i=1}^nr^2_i,$$$$
c=-\omega\sum_{i=1}^nr^3_i+(n+\beta-1)\sum_{i=1}^nr_i+\omega\sqrt{2h}+\frac{(n+\beta-1)}{\sqrt{2h}}\left(f-\frac{g^2}{4h}\right),$$$$
d=-\omega\sum_{i=1}^nr^4_i+(n+\beta-1)\sum_{i=1}^nr^2_i+\sum_{i<j}^nr_ir_j+\frac{1}{\sqrt{2h}}\left(f-\frac{g^2}{4h}\right)\sum_{i=1}^nr_i
\hspace{0.5in}$$$$\hspace{1in}+\frac{g(2n+2\beta-3)}{2\sqrt{2h}}+\frac{1}{4h}\left(f-\frac{g^2}{4h}\right)^2,
\end{equation}
where the roots $\{r_i\}$ satisfy the Bethe ansatz equations
\begin{equation}\label{eq:3.41}
\sum_{j\neq i}^n\frac{1}{r_i-r_j}+\frac{-\omega r_i^5+\beta r_i^3+\frac{r_i^2}{\sqrt{2h}}
\left(f-\frac{g^2}{4h}\right)+\frac{gr_i}{\sqrt{2h}}+\sqrt{2h}}{r_i^4}=0,~~~~i=1,2,\cdots, n.
\end{equation}

It can be  checked numerically using Maple that the wave functions $\Psi_n(r)$ are squarely integrable:
$\int^\infty_0\,|\Psi_n(r)|^2\,dr<\infty$.

As examples of the general expressions for the exact solutions, we consider the ground and first excited states of the system
in some detail. It is easy to see that $y=1$ is a solution of \eqref{eq:3.38} provided the potential parameters satisfy some constraints.
This solution corresponds to the $n=0$ case in the general expressions above. Indeed, from Eqs. \eqref{eq:2.12}, \eqref{parameter-beta},
\eqref{octic-wavefunction}, \eqref{eq:3.39} and \eqref{eq:3.40}, we obtain
\begin{equation}\label{eq:3.42}
E_0=\omega\left[\frac{5}{2}+\frac{e}{\sqrt{2h}}+\frac{g}{4}\sqrt{\frac{2}{h^3}}\left(\frac{g^2}{4h}-f\right)\right],
$$$$
\Psi_0(r)=r^\beta\exp\left(-\frac{\omega}{2}r^2-\frac{1}{\sqrt{2h}}\left(f-\frac{g^2}{4h}\right)\frac{1}{r}-\frac{g}{2\sqrt{2h}}\frac{1}{r^2}
-\frac{\sqrt{2h}}{3}\frac{1}{r^3}\right)
\end{equation}
with the constraints
\begin{equation}\label{eq:3.43}
a=-\frac{\omega}{\sqrt{2h}}\left(f-\frac{g^2}{4h}\right),$$$$
b=\frac{1}{2}\left[(\beta+\ell)(\beta-\ell-1)\right]-\frac{\omega g}{\sqrt{2h}},$$$$
c=-\omega\sqrt{2h}+\frac{(\beta-1)}{\sqrt{2h}}\left(f-\frac{g^2}{4h}\right),$$$$
d=\frac{1}{4h}\left(f-\frac{g^2}{4h}\right)^2+\frac{g(2\beta-3)}{2\sqrt{2h}}.
\end{equation}
The wave function \eqref{eq:3.42} has no nodes and so the state described by it is the ground state of the system.

The first excited state solution of the system corresponds to the $n=1$ case. From the general expressions above,
we obtain the first excited state energy and wave function,
\begin{equation}\label{eq:3.44}
E_1=\omega\left[\frac{7}{2}+\frac{e}{\sqrt{2h}}+\frac{g}{4}\sqrt{\frac{2}{h^3}}\left(\frac{g^2}{4h}-f\right)\right],
$$$$
\Psi_1(r)=r^\beta(r-r_1)\exp\left(-\frac{\omega}{2}r^2-\frac{1}{\sqrt{2h}}\left(f-\frac{g^2}{4h}\right)\frac{1}{r}
  -\frac{g}{2\sqrt{2h}}\frac{1}{r^2}-\frac{\sqrt{2h}}{3}\frac{1}{r^3}\right),
\end{equation}
where the root $r_1$ is determined by the Bethe ansatz equation,
\begin{equation}\label{eq:3.46}
-\omega r_1^5+\beta r_1^3+\frac{1}{\sqrt{2h}}\left(f-\frac{g^2}{4h}\right)r^2_1+\frac{g}{\sqrt{2h}}r_1+\sqrt{2h}=0
\end{equation}
and the potential parameters are constrained by
\begin{equation}\label{eq:3.45}
a=-\omega\left[\frac{1}{\sqrt{2h}}\left(f-\frac{g^2}{4h}\right)+r_1\right],$$$$
b=\frac{1}{2}\left[(\beta-\ell)(\beta+\ell+1)\right]+\frac{\omega g}{\sqrt{2h}}-\omega r_1^2,$$$$
c=-\omega r_1^3+\beta r_1-\omega\sqrt{2h}-\frac{\beta}{\sqrt{2h}}\left(f-\frac{g^2}{4h}\right),$$$$
d=-\omega r_1^4+\beta r_1^2+\frac{r_1}{\sqrt{2h}}\left(f-\frac{g^2}{4h}\right)
   +\frac{1}{4h}\left(f-\frac{g^2}{4h}\right)^2+\frac{g(2\beta-1)}{2\sqrt{2h}}.
\end{equation}

\vskip.2in
\noindent{\bf\large Case 2: $\omega=0$}
\vskip.1in
In this case, $A=-\omega/2=0$. Thus in order for the wave function $\Psi(r)$ to have acceptable asymptotic behaviour
at $r\rightarrow \infty$, we require the constant parameter $B$ to be negative.  Furthermore, we set
\begin{equation}
2E+B^2=0.
\end{equation}
Then equation \eqref{eq:2.14} becomes
\begin{equation}\label{eq:3.47}
r^4y''(r)+2\left[Br^4+\beta r^3+\frac{r^2}{\sqrt{2h}}\left(f-\frac{g^2}{4h}\right)+\frac{gr}{\sqrt{2h}}+\sqrt{2h}\right]y'(r)$$$$
 +\left\{2\left[\beta B-a\right]r^3+\left[(\beta-\ell)(\beta-\ell-1)-2b+\frac{2B}{\sqrt{2h}}\left(f-\frac{g^2}{4h}\right)\right]r^2\right.
$$$$\left.-2\left[c-\frac{\beta-1}{\sqrt{2h}}\left(f-\frac{g^2}{4h}\right)-\frac{gB}{\sqrt{2h}}\right]r\right\}y(r)
$$$$\hspace{0.5in}=\left[2d-\frac{g(2\beta-3)}{\sqrt{2h}}-2B\sqrt{2h}-\frac{1}{2h}\left(f-\frac{g^2}{4h}\right)^2\right]y(r).
\end{equation}
Substituting \eqref{octic-wavefunction} into this equation and applying the results in the results of the Appendix,
we obtain constraints for the potential parameters,
\begin{equation}\label{eq:3.49}
B=\frac{a}{n+\beta},$$$$
b=\frac{1}{2}\left[(\beta+\ell)(\beta-\ell-1)+n(n+2\beta-1)\right]+B\left[\frac{1}{\sqrt{2h}}\left(f-\frac{g^2}{4h}\right)+\sum_{i=1}^nr_i\right],
$$$$
c=B\left[\frac{g}{\sqrt{2h}}+\sum_{i=1}^nr_i^2\right]+(n+\beta-1)\left[\frac{1}{\sqrt{2h}}\left(f-\frac{g^2}{4h}\right)+\sum_{i=1}^nr_i\right],
$$$$
d=B\sum_{i=1}^nr_i^3+(n+\beta-1)\sum_{i=1}^nr_i^2+\sum_{i<j}^nr_ir_j+\frac{1}{\sqrt{2h}}\left(f-\frac{g^2}{4h}\right)\sum_{i=1}^nr_i
$$$$+\frac{1}{4h}\left(f-\frac{g^2}{4h}\right)^2+B\sqrt{2h}+\frac{g(2n+2\beta-3)}{2\sqrt{2h}},
\end{equation}
where $\beta$ is given by Eq. \eqref{parameter-beta} and the roots $\{r_i\}$ satisfy the Bethe ansatz equations
\begin{equation}\label{eq:3.50}
\sum_{j\neq i}^n\frac{1}{r_i-r_j}+\frac{B r_i^4+\beta r_i^3+\frac{r_i^2}{\sqrt{2h}}\left(f-\frac{g^2}{4h}\right)
+\frac{gr_i}{\sqrt{2h}}+\sqrt{2h}}{r_i^4}=0,~~~~n=1,2,\cdots,n.
\end{equation}
It follows that the energies and the wave functions of the system are
\begin{equation}\label{eq:3.48}
E_n=-\frac{1}{2}B^2=-\frac{1}{2}\left(\frac{a}{n+\beta}\right)^2,$$$$
\Psi_n(r)=r^\beta \left[\prod_{i=1}^n(r-r_i)\right]\exp\left(\frac{a}{n+\beta}r-\frac{1}{\sqrt{2h}}
\left(f-\frac{g^2}{4h}\right)\frac{1}{r}-\frac{g}{2\sqrt{2h}}\frac{1}{r^2}-\frac{\sqrt{2h}}{3}\frac{1}{r^3}\right).
\end{equation}

As mentioned above, $B$ is require to be negative. To satisfy this requirement we see from \eqref{eq:3.49}
that the potential parameter $a$ has to be negative. So we assume that $a<0$ if $\omega=0$.
Then it can be  checked numerically that the wave functions $\Psi_n(r)$ are squarely integrable.

Similar to the Case 1 above, the ground state solution is obtained from the general expressions by letting $n=0$, yielding
\begin{equation}\label{eq:3.51}
E_0=-\frac{a^2}{2\beta^2},
$$$$
\Psi_0(r)=r^\beta\exp\left(\frac{a}{\beta}r-\frac{1}{\sqrt{2h}}\left(f-\frac{g^2}{4h}\right)\frac{1}{r}-\frac{g}{2\sqrt{2h}}\frac{1}{r^2}
-\frac{\sqrt{2h}}{3}\frac{1}{r^3}\right)
\end{equation}
with the constraints
\begin{equation}\label{eq:3.52}
b=\frac{1}{2}\left[\beta(\beta-1)-\ell(\ell+1)\right]+\frac{a}{\beta\sqrt{2h}}\left(f-\frac{g^2}{4h}\right),
$$$$c=\frac{ag}{\beta\sqrt{2h}}+\frac{(\beta-1)}{\sqrt{2h}}\left(f-\frac{g^2}{4h}\right),
$$$$d=\frac{1}{{4h}}\left(f-\frac{g^2}{4h}\right)^2+\frac{a\sqrt{2h}}{\beta}+\frac{g(2\beta-3)}{2\sqrt{2h}}.
\end{equation}
The first excited state solution corresponds to the $n=1$ case. We thus have
\begin{equation}\label{eq:3.53}
E_1=-\frac{1}{2}\left(\frac{a}{1+\beta}\right)^2,$$$$
\Psi_1(r)=r^\beta(r-r_1)\exp\left(\frac{a}{1+\beta}r-\frac{1}{\sqrt{2h}}\left(f-\frac{g^2}{4h}\right)\frac{1}{r}
   -\frac{g}{2\sqrt{2h}}\frac{1}{r^2}-\frac{\sqrt{2h}}{3}\frac{1}{r^3}\right)
\end{equation}
subject to the constraints
\begin{equation}\label{eq:3.54}
b=\frac{1}{2}(\beta-\ell)(\beta+\ell+1)+\frac{a}{1+\beta}\left[\frac{1}{\sqrt{2h}}\left(f-\frac{g^2}{4h}\right)+r_1\right],
$$$$c=\frac{a}{1+\beta}r_1^2+\beta r_1+\frac{ag}{(1+\beta)\sqrt{2h}}+\frac{\beta}{\sqrt{2h}}\left(f-\frac{g^2}{4h}\right),
$$$$d=\frac{a}{1+\beta}r_1^3+\beta r_1^2+\frac{r_1}{\sqrt{2h}}\left(f-\frac{g^2}{4h}\right)
  +\frac{1}{4h}\left(f-\frac{g^2}{4h}\right)^2+\frac{a\sqrt{2h}}{\beta+1}+\frac{g(2\beta-1)}{2\sqrt{2h}}.
\end{equation}
Here the root $r_1$ obeys
\begin{equation}\label{eq:3.54b}
\frac{a}{1+\beta}r_1^4+\beta r_1^3+\frac{1}{\sqrt{2h}}\left(f-\frac{g^2}{4h}\right)r_1^2+\frac{g}{\sqrt{2h}}r_1+\sqrt{2h}=0.
\end{equation}

Let us remark that the results obtained in this section are reducible to those of the inverse quartic potential when
$g,f,e,h\rightarrow 0$ so that $\beta\rightarrow 1+c/\sqrt{2d}$ and $\frac{1}{\sqrt{2h}}\left(f-\frac{g^2}{4h}\right)\rightarrow \sqrt{2d}$,
and reproduce those of the inverse sextic potential if $a,b,c,e,g,h\rightarrow 0$ such that
$\beta\rightarrow 1+d/\sqrt{2f}$ and $g/\sqrt{2h}\rightarrow \sqrt{2f}$.

\section{Quasi-exactly solvable inverse decatic power potential }
We consider the inverse decatic power potential
\begin{equation}\label{eq:2.15}
V(r)=\frac{a}{r^4}+\frac{b}{r^6}+\frac{c}{r^8}+\frac{d}{r^{10}},
\end{equation}
where $d >0$. The corresponding radial Schr\"odinger equation is given by
\begin{equation}
\left[-\frac{d^2}{dr^2}+\frac{\ell(\ell+1)}{r^2}+\omega^2r^2+\frac{2a}{r^4}+\frac{2b}{r^6}+\frac{2c}{r^8}
+\frac{2d}{r^{10}}\right]\Psi(r).
=2E\Psi(r)
\end{equation}
Similar to the cases studied in the previous sections, finding exact solutions to this equation requires
first extracting appropriate asymptotic behaviour of the wave function $\Psi(r)$. This is achieved by means of
suitable transformations of $\Psi(r)$. After some algebras we arrive at the transformation
\begin{equation}\label{eq:2.16}
\Psi(r)=r^\eta\exp\left(-\frac{\omega}{2}r^2-\frac{c}{2\sqrt{2d}}\frac{1}{r^2}-\frac{\sqrt{2d}}{4}\frac{1}{r^4}\right)u(r),
$$$$ \eta=\frac{5}{2}+\frac{b}{\sqrt{2d}}+\frac{c^2}{16}\sqrt{\frac{2}{d^3}}>0.
\end{equation}
Then the differential equation for $u(r)$ is
\begin{equation}\label{eq:2.18}
u''(r)+\frac{2}{r}\left(-\frac{\sqrt{2d}}{r^4}+\frac{c}{r^2\sqrt{2d}}+\eta-\omega r^2\right)u'(r)
\hspace{1in}$$$$+\left[-\omega(\eta+1/2)+\frac{\eta(\eta-1)-\ell(\ell+1)-2\omega c/\sqrt{2d}}{r^2}
-\frac{{3c}+2d\omega-2c\eta +2a\sqrt{2d}}{\sqrt{2d}\,r^4}\right]u(r)\hspace{1in}$$$$+2Eu(r)=0.
\end{equation}
A further variable change $z=r^2$ transforms the above equation into the form
\begin{equation}\label{eq:2.19}
z^3u''(z)+\left[-\omega z^3+\left(\eta+1/2\right)z^2+\frac{c}{\sqrt{2d}}z+\sqrt{2d}\right]u'(z)+\left\{\frac{1}{2}
\left[E-\omega(\eta+1/2)\right]z^2\right.$$$$\left.+\frac{z}{4}\left[\eta(\eta-1)-\ell(\ell+1)-\frac{2\omega c}{\sqrt{2d}}\right]\right\}u(z)
=\frac{1}{2}\left[{a}+\omega\sqrt{2d}+\frac{c}{\sqrt{2d}}(3/2-\eta)\right]u(z).
\end{equation}

This equation is quasi-exactly solvable and has the degree $n$ polynomial solutions
\begin{equation}\label{decatic-wavefunction}
u(z)=\prod_{i=1}^n(z-z_i),\hspace{0.3in}u(z)\equiv 1\hspace{0.1in}\mbox{for}\hspace{0.1in} n=0
\end{equation}
with distinct roots $z_i$ provided that the potential parameters satisfy certain constraints. Indeed,
substituting \eqref{decatic-wavefunction} into \eqref{eq:2.19} and applying the functional Bethe ansatz method
in \cite{Zhang11} outlined in the Appendix, we get the closed form expressions for the energies and the wave functions
\begin{equation}\label{eq:3.55}
E_n=\omega\left(2n+\eta+\frac{1}{2}\right)=\omega\left(2n+3+\frac{b}{\sqrt{2d}}+\frac{c^2}{16}\sqrt{\frac{2}{d^3}}\right),$$$$
\Psi_n(r)=r^\eta\left[\prod_{i=1}^n(r^2-z_i)\right]\exp\left(-\frac{\omega}{2}r^2-\frac{c}{2\sqrt{2d}}\frac{1}{r^2}-\frac{\sqrt{2d}}{4}\frac{1}{r^4}\right)
\end{equation}
and the constraints for the potential parameters
\begin{equation}\label{eq:3.56}
2\omega\left(\frac{c}{\sqrt{2d}}+2\sum_{i=1}^nz_i\right)=\left(\frac{c^2}{16}\sqrt{\frac{2}{d^3}}+2\right)^2-\left(\ell+\frac{1}{2}\right)^2
+4n\left(n+2+\frac{c^2}{16}\sqrt{\frac{2}{d^3}}\right),
$$$$
a=-2\omega\sum_{i=1}^nz_i^2+\left(4n+2+\frac{c^2}{8}\sqrt{\frac{2}{d^3}}\right)\sum_{i=1}^nz_i
+\frac{c}{\sqrt{2d}}(2n+1)+\frac{c^3}{16d^2}-\omega\sqrt{2d},
\end{equation}
where the roots $\{z_i\}$ obey the Bethe ansatz equations
\begin{equation}\label{eq:3.57}
\sum_{j\neq i}^n\frac{1}{z_i-z_j}+\frac{-\omega\sqrt{2d}~z^3_i+\left(3\sqrt{2d}+c^2/8d\right)z_i^2+cz_i+2d}{\sqrt{2d}~z^3_i}=0,~~~~
i=1,2,\cdots,n.
\end{equation}
The wave functions $\Psi_n(r)$ are squarely integrable, i.e. $\int^\infty_0\,|\Psi_n(r)|^2\,dr<\infty$.
This can be checked numerically using Maple.

Clearly, the ground state solution can be obtained from the above general expressions by letting $n=0$, yielding
\begin{equation}\label{decatic-E0}
E_0=\omega\left(3+\frac{b}{\sqrt{2d}}+\frac{c^2}{16}\sqrt{\frac{2}{d^3}}\right),$$$$
\Psi_0(r)=r^{\frac{5}{2}+\frac{b}{\sqrt{2d}}+\frac{c^2}{16}\sqrt{\frac{2}{d^3}}}\exp\left(-\frac{\omega}{2}r^2-\frac{c}{2\sqrt{2d}}\frac{1}{r^2}
-\frac{\sqrt{2d}}{4}\frac{1}{r^4}\right)
\end{equation}
with the constraints
\begin{equation}\label{eq:3.58}
\frac{2c}{\sqrt{2d}}\omega=\left(\frac{c^2}{16}\sqrt{\frac{2}{d^3}}+2\right)^2-\left(\ell+\frac{1}{2}\right)^2,$$$$
a= \frac{c}{\sqrt{2d}}+\frac{c^3}{16d^2}-\omega\sqrt{2d}.
\end{equation}
The first excited state solution corresponds to the $n=1$ case. We thus have the energy and the wave function,
\begin{equation}\label{decatic-E1}
E_1=\omega\left(5+\frac{b}{\sqrt{2d}}+\frac{c^2}{16}\sqrt{\frac{2}{d^3}}\right),$$$$
\Psi_1(r)=r^{{\frac{5}{2}+\frac{b}{\sqrt{2d}}+\frac{c^2}{16}\sqrt{\frac{2}{d^3}}}}(r^2-z_1)\exp\left(-\frac{\omega}{2}r^2-\frac{c}{2\sqrt{2d}}
\frac{1}{r^2}-\frac{\sqrt{2d}}{4}\frac{1}{r^4}\right)
\end{equation}
with the potential parameters subject to the constraints
\begin{equation}\label{eq:3.59}
\omega\left(c+2\sqrt{2d}z_1\right)=\frac{\sqrt{2d}}{2}\left[\left(\frac{c^2}{16}\sqrt{\frac{2}{d^3}}
+2\right)^2-\left(\ell+\frac{1}{2}\right)^2\right]+6\sqrt{2d}+\frac{c^2}{4d},
$$$$
a=-2\omega z_1^2+\left(6+\frac{c^2}{8}\sqrt{\frac{2}{d^3}}\right)z_1 +\frac{3c}{\sqrt{2d}}+\frac{c^3}{16d^2}-\omega\sqrt{2d}.
\end{equation}
Here the root $z_1$ is determined by the Bethe ansatz equation,
\begin{equation}\label{eq:3.60}
-\omega\sqrt{2d}~z^3_1+\left(3\sqrt{2d}+c^2/8d\right)z_1^2+cz_1+2d=0.
\end{equation}

\section{Concluding remarks}
We have presented four classes of quasi-exactly solvable anharmonic singular potentials and solved the corresponding
radial Schr\"odinger equations exactly by means of the Bethe ansatz method. Closed-form expressions for the energies and
the wave functions of the singular models have been obtained for certain values of the potential parameters.
Analytic expressions for the allowed potential parameters are given explicitly for each of the four cases
in terms of the roots of a set of algebraic Bethe ansatz equations.

It is remarkable that the four classes of anharmonic singular potentials are quasi-exactly solvable. We have thus added
four new members to the club of quasi-exactly solvable quantum mechanical systems. It is hoped that our findings in
the present paper will lead to new applications for these singular potentials.

\section*{Acknowledgments }
This work was supported in part by the Australian Research Council through Discovery Project DP110103434.
D. Agboola acknowledges the support of an International Postgraduate Research Scholarship
and a University of Queensland Centennial Scholarship. Y.-Z. Zhang acknowledges
the partial support of an Alexander von Humboldt Fellowship and thank the Physikalisches Institut der Universit\"at Bonn,
especially G\"unter von Gehlen, for hospitality.


\appendix

\section{Appendix: the functional Bethe ansatz method}
As seen from \eqref{eq:3.38} and \eqref{eq:3.47},
the inverse octic power potential leads to the 2nd order differential equations of the form,
\begin{equation}\label{eq:3.1}
\left[P(t)\frac{d^2}{dt^2}+Q(t)\frac{d}{dt}+W(t)\right]S(t)=0,
\end{equation}
where $P(t), Q(t)$ and $W(t)$ are polynomials of at most degree 4, 5 and 4 respectively,
\begin{equation}\label{eq:3.2}
P(t)=\sum_{k=0}^4p_kt^k,\hspace{0.2in}Q(t)=\sum_{k=0}^5q_kt^k,\hspace{0.2in}W(t)=\sum_{k=0}^4w_kt^k,
\end{equation}
$p_k, q_k$ and $w_k$ are constants. This 2nd order differential equation is a generalization of the one considered
in \cite{Zhang11} and does not seem to have ever appeared previously in the literature. A similar differential
equation also appears in quantum mechanical models with (non-singular) octic potentials \cite{AZ12}.

In this appendix, we show that the differential equation \eqref{eq:3.1} is quasi-exactly solvable
for certain values of its parameters and exact solutions are given
by polynomials of degree $n$ in $t$ with $n$ being non-negative integers.
We apply the functional Bethe ansatz method in \cite{Zhang11}. We seek polynomial solutions of the form
\begin{equation}\label{eq:3.3}
S(t)=\prod_{i=1}^n(t-t_i),\hspace{0.3in}S(t)\equiv 1\hspace{0.1in}\mbox{for}\hspace{0.1in} n=0
\end{equation}
with $t_1, t_2,\dots,t_n$ distinct roots to be determined, then  Eq.\,\eqref{eq:3.1} becomes
\begin{eqnarray}
-w_0&=&\left(p_4t^4+p_3t^3+p_2t^2+p_1t+p_0\right)\sum_{i=1}^n\frac{1}{t-t_i}\sum_{j\neq i}^n\frac{2}{t_i-t_j}\nonumber\\
& &+\left(q_5t^5+q_4t^4+q_3t^3+q_2t^2+q_1t+q_0\right)\sum_{i=1}^n\frac{1}{t-t_i}\nonumber\\
& &+w_4t^4+w_3t^3+w_2t^2+w_1t\label{eq:3.4}
\end{eqnarray}
The left hand side of this equation is a constant, while the right hand side is a meromorphic
function with simple poles $t=t_i$ and singularity at $t=\infty$. For this equation to be valid,
the right hand side must also be a constant. We thus demand that the coefficients of the powers of
$t$ as well as the residues at the simple poles of the right hand side be zero.
Following Liouville's theorem, this is the necessary and sufficient condition  for the right hand side
of \eqref{eq:3.4} to be a constant.  It follows that
\begin{eqnarray}
w_4&=&-nq_5,\label{w4-formula}\\
w_3&=&-q_5 \sum_{i=1}^nt_i-nq_4,\\
w_2&=&-q_5\sum_{i=1}^nt_i^2-q_4\sum_{i=1}^nt_i-n(n-1)p_4-nq_3,\\
w_1&=&-q_5\sum_{i=1}^nt^3_i-q_4\sum_{i=1}^nt^2_i-\left[2(n-1)p_4+q_3\right]\sum_{i=1}^nt_i-n(n-1)p_3-nq_2,\\
w_0&=&-q_5\sum_{i=1}^nt^4_i-q_4\sum_{i=1}^nt_i^3-\left[q_3+2(n-1)p_4\right]\sum_{i=1}^nt_i^2-2p_4\sum_{i<j}^nt_it_j\nonumber\\
& &-\left[2(n-1)p_3+q_2\right]\sum_{i=1}^nt_i-n(n-1)p_2-nq_1,\label{w0-formula}
\end{eqnarray}
where the roots $t_1,t_2,\dots,t_n$ satisfy the Bethe ansatz equations
\begin{equation}\label{eq:3.9}
\sum_{j\neq i}^n\frac{2}{t_i-t_j}+\frac{q_5t_i^5+q_4t_i^4+q_3t_i^3+q_2t_i^2+q_1t_i+q_0}{p_4t_i^4+p_3t_i^3+p_2t_i^2+p_1t_i+p_0}=0,
\hspace{0.2in}i=1,2,\dots,n
\end{equation}
Equations \eqref{w4-formula}-\eqref{w0-formula} give all polynomials $W(t)$ such that \eqref{eq:3.1}
has degree $n$ polynomial solutions \eqref{eq:3.3}. The above method can be easily
generalized to differential equations with higher degree polynomials $P(t)$, $Q(t)$ and $W(t)$, provided that $\deg W(t)<\deg Q(t)$.


\end{document}